\documentclass{article}
\usepackage{frascatiphys}
\usepackage{subfigure}
\usepackage[dvips]{pstcol}
\usepackage[dvips]{graphicx}
\usepackage{amssymb}
\usepackage{bm}%
\begin{document}
\title{ 
CHALLENGES IN HADRON PHYSICS
}
\author{
Kamal K. Seth\\
\textit{Northwestern University, Evanston, IL 60208, USA}\\
kseth@northwestern.edu
}

\maketitle
\baselineskip=11.6pt
\begin{abstract}
The status of hadron physics at the end of the HADRON07 Conference is reviewed.  The latest results presented at the conference, as well as those important developments in the field which were not represented, are included.
\end{abstract}
\baselineskip=14pt

For this closing talk, I was encouraged to be different and not attempt to mention over and over again the talks that you have already heard.  And there have been a lot of talks, 36 plenaries and 146 parallel presentations.  Instead, I will try to present a summary of the challenges we face at the end of 2007 as hadron spectroscopists, what we have achieved and what we must strive to achieve.

This series of two--yearly HADRON conferences began at Maryland in 1983.  The first one I attended in 1991 defined our charter as \textbf{``hadron spectroscopy and some areas of related hadron structure''}, i.e., strong interaction physics, which in the modern language means QCD. So, let me walk across the landscape of hadron physics and survey the challenges.  I will talk about things which have been presented at this conference, as well as those which have not.  And the presentation will be admittedly subjective.

For an overall survey of the recent progress in experimental hadron spectroscopy, I will often turn to the PDG, which provides us with the only Bible we have, imperfect as it might be.

\section{BARYONS}

Two quarks are easier than three, but I begin with baryons because we live in a Universe built of baryons, to be more specific --- nucleons, and not mesons.

\subsection{The Nucleons}

We have been working on the nucleon for close to 100 years, and all we want to know is what the nucleon looks like.  How do its static properties, mass, charge, magnetic moment, spin, and structure arise, and how it reacts when it is tickled by an external probe?  Not too much to ask!!

We are told that there is a super-duper new way of tickling the nucleon, \textbf{Deeply Virtual Compton Scattering (DVCS)/Deeply Virtual Meson Production (DVMP)}, which leads to the Generalized Parton Distributions (GPD's), and they can give all the information that we used to try to get by measuring form factors and Deep Inelastic Scattering (DIS).  Maybe so!

Indeed, nearly all the labs in the business, JLab, H1, Zeus, Hermes, Brookhaven, CERN(COMPASS), now have very active programs for measuring GPD's (see Fig.~1).

\begin{figure}[!tb]
\begin{center}
\includegraphics[width=2.3in]{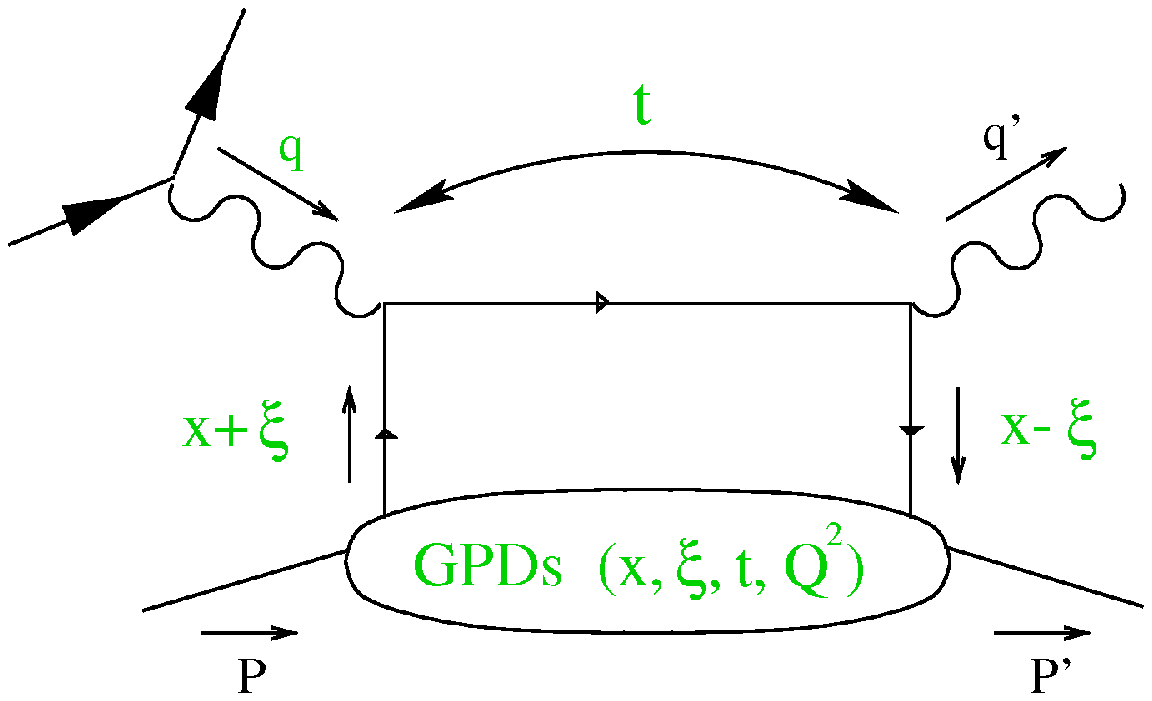}
\includegraphics[width=2.3in]{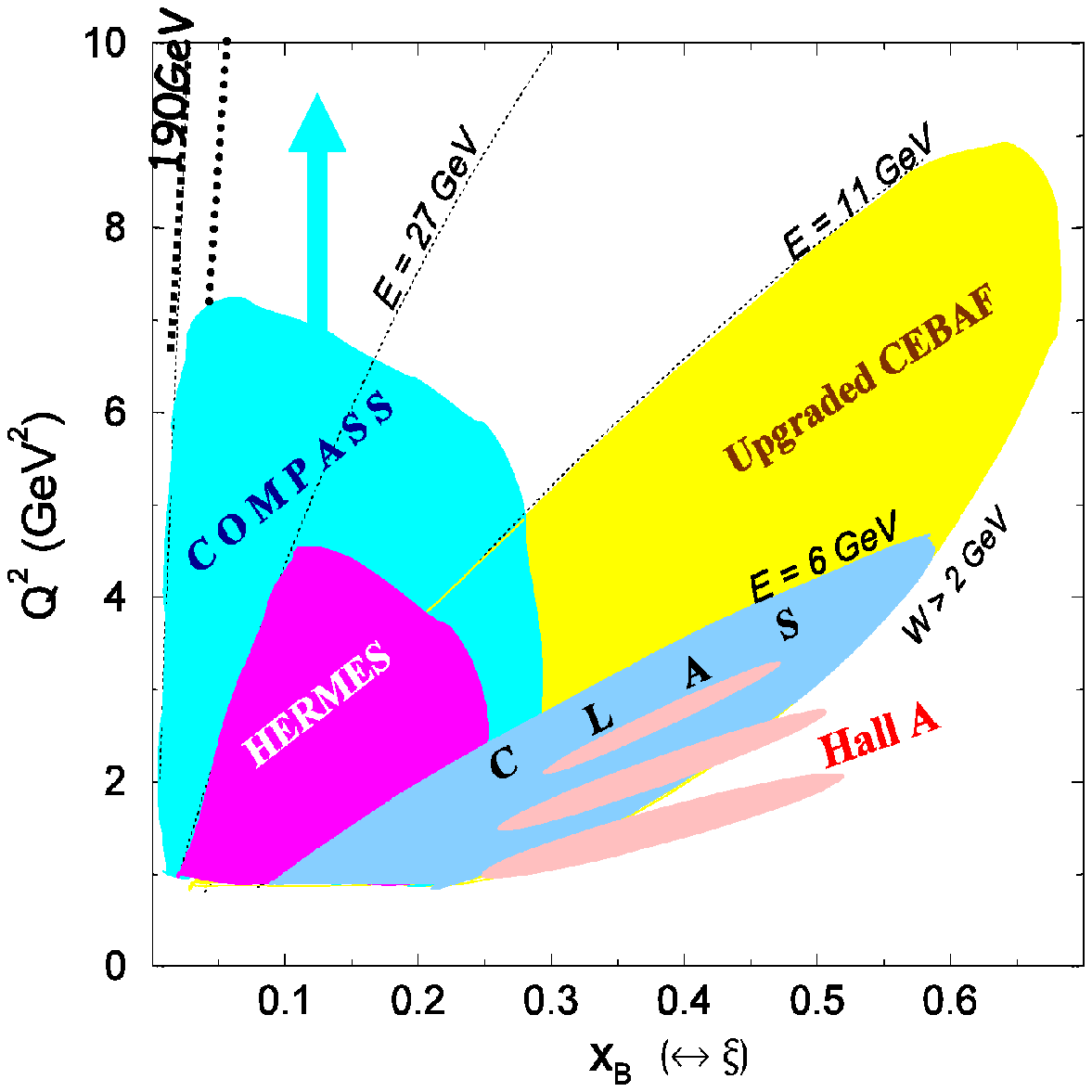}
\end{center}
\caption{(Left) Schematic representation of the DVCS measurement of GPD's.  (Right) Illustrating the $x_B-Q^2$ domain accessible to GPD measurements at different laboratories.}
\end{figure}

Unfortunately, life is not so easy.  The observables are all integrals over $x$(Bjorken) and deconvolution is required to get to the true GPD's, $H,~\tilde{H},~E,~\tilde{E}$, which are functions  of $x$, $\xi$, and $t$.  And that is neither easy nor unique.  So, like the GDP, which is supposed to contain all the information about a country's economy, the whole story is not in the GPD measurements, either. For the present, therefore, we go back to the more directly interpretable form factor and DIS measurements.

\subsection{Form Factors}

There has been recent progress in form factor measurements, both for spacelike and timelike momentum transfers.\vspace{8pt}

\noindent\textbf{1. The Challenge of $\bm{G_E}$(proton) for Spacelike Momentum Transfers}\vspace{6pt}

The recent JLab polarization measurements for spacelike momentum transfers up to $Q^2=5.5$~GeV$^2$ show clearly that the $\mu G_E(p)/G_M(p)$ of the proton monotonically falls with increasing $Q^2$, or equivalently, $Q^2F_2/F_1$ monotonically rises, as shown in Fig.~2.  The naive pQCD expectation was that both $\mu G_E(p)/G_M(p)$ and $Q^2F_2/F_1$ were constant for large $Q^2$.  What happened?  Many \textbf{post}dictions have been made.  Suffice it to say that there are no clear-cut consensus explanations.  JLab proposes to extend these measurements to $Q^2=8.6$~GeV$^2$ by which time $\mu G_E(p)/G_M(p)$ should have arrived at zero, i.e., $G_E=0$.  What does $G_E=0$, mean? If the trend continues, with the 12 GeV upgrade $G_E$ will be found to be negative.  What does \textbf{that} mean?

We do not know, but it is clear that the measurements have to be made, and the theorists have to work harder to tell us what they mean.\vspace{30pt}

\begin{figure}[!tb]
\begin{center}
\includegraphics[width=2.3in]{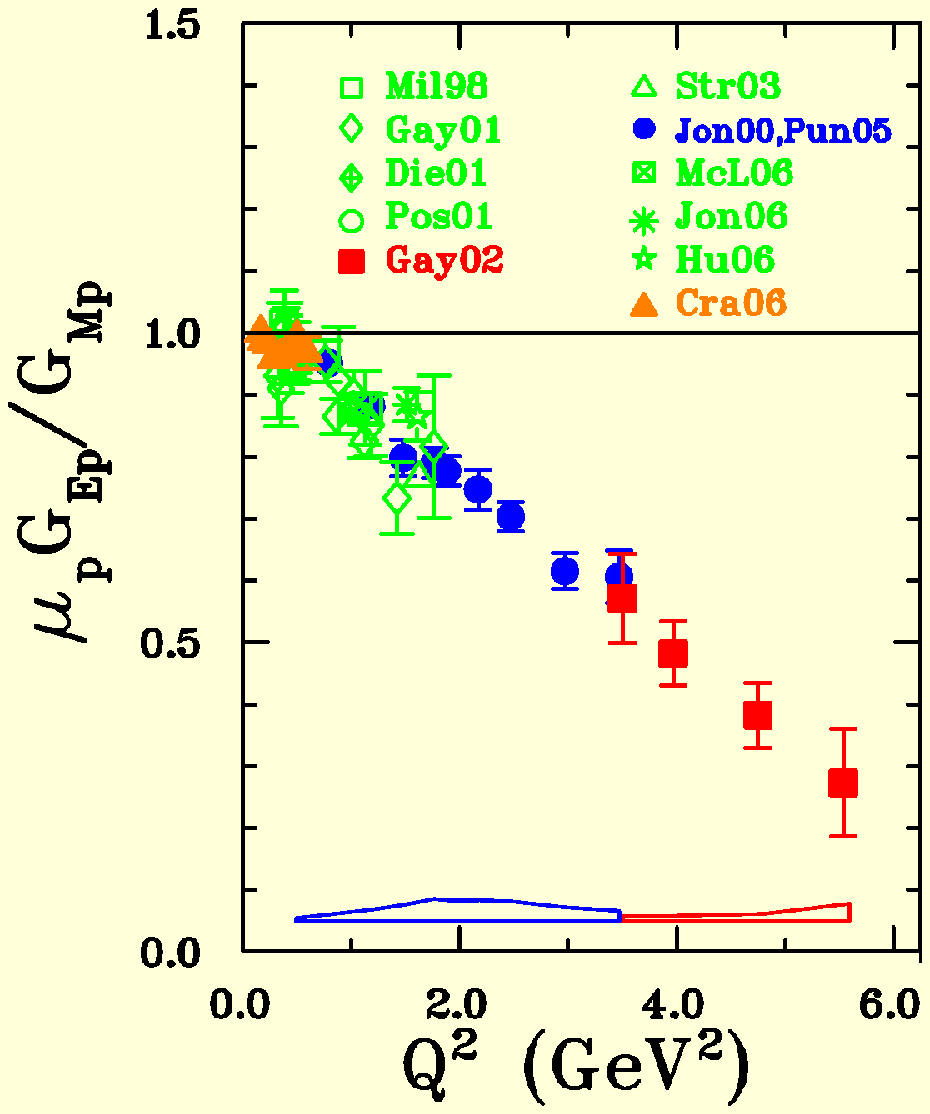}
\includegraphics[width=2.3in]{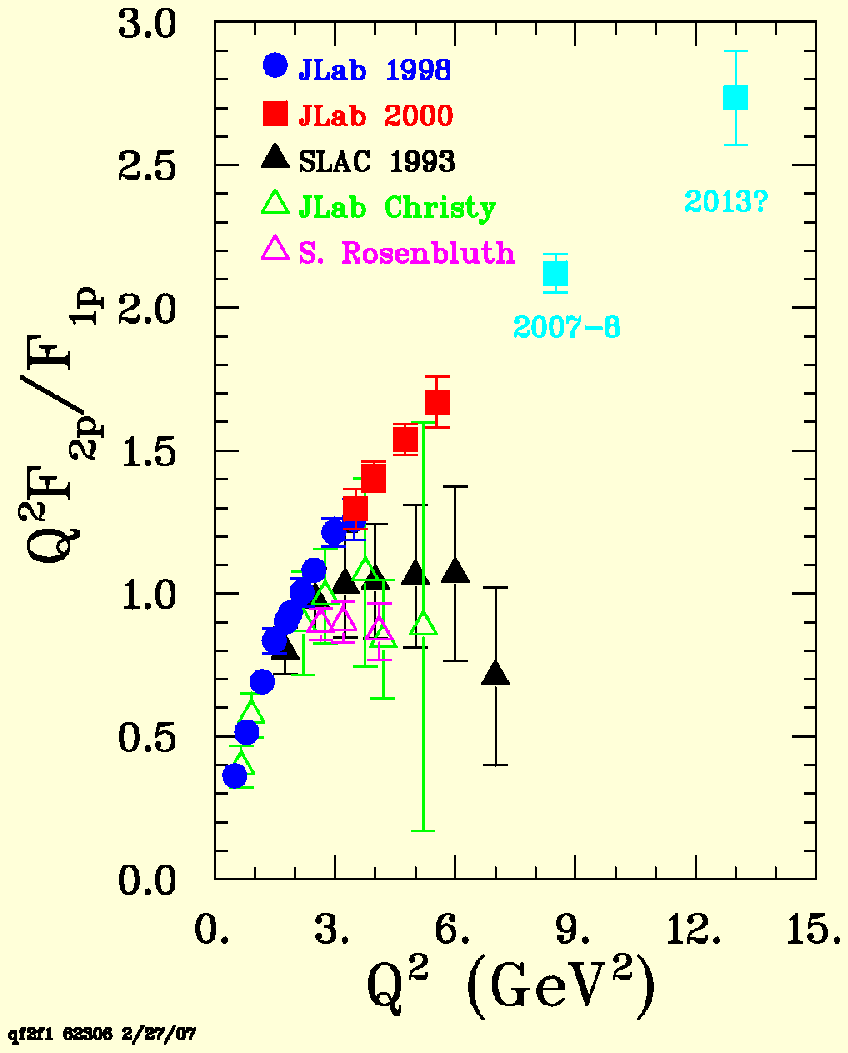}
\end{center}
\caption{(Left) Results for $\mu G^p_E/G^p_M$ as a function of $Q^2$ as measured in the polarization experiments at JLab.  (Right) The results presented as $Q^2F^p_2/F^p_1$.}
\end{figure}

\noindent\textbf{2. The Challenge of $\bm{G_E}$(proton) for Timelike Momentum Transfers}\vspace{6pt}

In the perturbative regime of large momentum transfer $Q^2$, QCD makes two predictions about the relationship between form factors at spacelike and timelike momentum transfers.  The first, based on quark counting rules, is that for both, $Q^4G_M(p)$ should be constant except for the variation of $\alpha_S^2$.  The second is that in this regime, the two should be equal (actually, this follows from Cauchy's theorem because form factors are analytic functions).  The general expectation was that $Q^2\ge10$~GeV$^2$ should be large enough for both these predictions to be true.  Fermilab $p\bar{p}\to e^+e^-$ measurements for $Q^2=8.8-13.1$~GeV$^2$ showed that while the $1/Q^4$ and $\alpha_S^2$ variations were essentially confirmed, the timelike form factors were twice as large as the corresponding spacelike form factors (see~Fig.~3).  Since then, the experimental measurements of Fermilab have been confirmed and extended by the reverse reaction measurements of $e^+e^-\to p\bar{p}$ by BES, CLEO, and BaBar.  So, the factor two is more than confirmed, and we have to understand how it arises. Let me add a very important point here.  For timelike form factors, lattice is totally impotant.  The practitioners admit that they are stuck in Euclidean time and cannot handle timelike form factors because they live in Minkowski time.  So, there!

This poses a challenge.  Why half--agreement and half--disagreement with pQCD?  One way out that has been suggested is that the proton does not look like a Mercedes star, with three symmetrically placed quarks, but more like a T, with a diquark-quark configuration, and the diquark model does succeed in explaining the factor two discrepancy.  Many people do not buy the diquark model, and so seek refuge in the possibility that $Q^2\sim15$~GeV$^2$ is not large enough for pQCD to be valid.  This, of course, throws the challenge to the experimentalists:  measure timelike form factors at larger $Q^2$.  Easier said than done!!  Recall that $G^p_M(|Q^2|)$ varies as $1/Q^4$, and the cross section varies as $1/Q^{10}$ or $1/s^5$. So, in going from 15~GeV$^2$ to 25~GeV$^2$ the cross section would fall a factor 20 from $<1$~pb to \textbf{$\bm{<50}$~fb}.  That is a difficult measurement. In principle, BES III and PANDA could tackle it, but it will be very hard.

\begin{figure}[!tb]
\begin{center}
\includegraphics[width=3.5in]{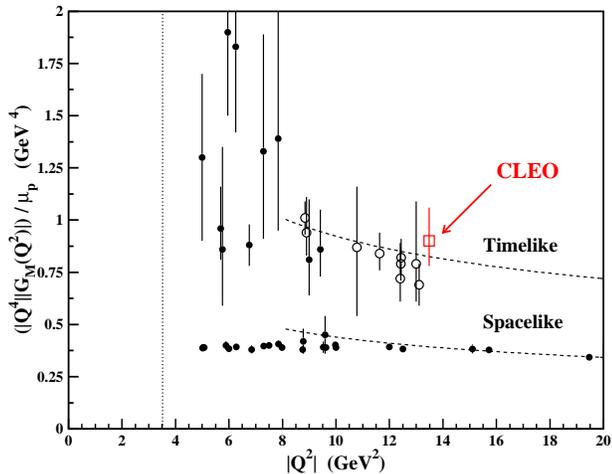}
\end{center}
\caption{Measurements of timelike magnetic form factors of the proton presented as $Q^4G_M/\mu_p$ versus $|Q^2|$.  Also shown are the spacelike form factors as measured at SLAC.  The dotted curves illustrate the $\alpha_S^2$ variation predicted by pQCD.}
\end{figure}

What about $G_E(p)/G_M(p)$ for timelike momentum transfers?  In principle, this could be done because the angular dependence of $G_E$ and $G_M$ is different.  In fact, BaBar has tried to do this in their measurement of $p\bar{p}$ production in ISR--mediated annihilation of $e^+e^-$.  The errors are large (and there is the familiar Rosenbluth ansatz), but essentially $\mu_p G_E/G_M$ is found to be constant $\approx1.3\pm0.2$ in the entire region $Q^2=3.5-9.0$~GeV$^2$.  Recall that by $Q^2=5.4$~GeV$^2$ the spacelike $\mu_pG_E/G_M$ has fallen down to $\sim0.3$, and it extrapolates to zero by $Q^2\approx 9$~GeV$^2$.  If the BaBar results hold up with better statistics, we have a serious problem on our hands.  To confirm and reconcile these results is an important challenge to both theorists and experimentalists.\vspace{8pt}

\begin{figure}[!tb]
\begin{center}
\includegraphics[width=4.in]{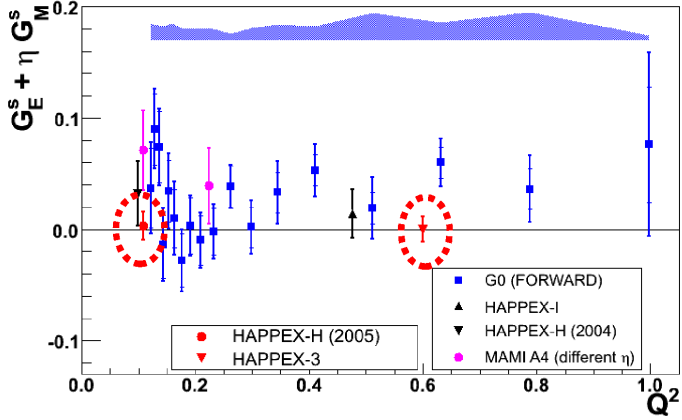}
\end{center}
\caption{Strange quark form factors ($G_E+\eta G_M$) as measured by parity violating electron scattering.}
\end{figure}

\noindent\textbf{3. The Challenge of Strange Quark Form Factors}\vspace{6pt}

For a long time there has been the nagging question about the role of the strange quarks in the nucleon.  Several experiments (SAMPLE at Bates(MIT), PVA4 at Mainz, and G$^0$ and HAPPEX at JLab) have addressed this question by making the very demanding measurements of parity violating electron scattering.  So far the data have been limited to $Q^2\le1$~GeV$^2$.

A global analysis of the world data for $Q^2\le0.48$~GeV$^2$ leads to the conclusion that for $Q^2=0.1$~GeV$^2$, $G^s_E(p)=-0.008\pm0.016$, and $G^s_M(p)=0.29\pm0.21$, i.e., both are consistent with zero.  The same analysis reaches essentially the same conclusion for the sixteen individual $Q^2$ data points from 0.12--0.48~GeV$^2$ (see~Fig.~4).

The experimental challenge is to see if this conclusion holds for larger $Q^2$ at which both G$^0$ and HAPPEX plan to take data in the near future.

Before I leave form factors, let me add that there has been very encouraging progress recently in measuring form factors of the neutron.  Excellent JLab measurements of $G_M^{(n)}$ extend up to $\sim4.7$~GeV$^2$ and for $G_E^{(n)}$ up to $\sim1.5$~GeV$^2$, and there are plans to go to larger $Q^2$.

\subsection{The Challenge of the Nucleon Spin}

We all know what this is about.  The quark spins just don't add up to the spin~1/2 of the proton.  So what accounts for the rest?

Proton spin$~=~1/2~=~\frac{1}{2}\Delta \Sigma +\Delta G + L_z$,

where $\Delta \Sigma = \Delta u + \Delta d + \Delta s$,\qquad $\Delta q = (q_+ - q_-) + (\bar{q}_+ - \bar{q}_-)$

The latest results are
\begin{center}
$\Delta \Sigma = 0.35\pm0.06$~(COMPASS), $0.33\pm0.04$~(HERMES)
\end{center}

That leaves a large part for $(\Delta G + L_z)$ to account for.

Attempts have been made to measure $\Delta G$ via DIS, polarized semi-inclusive DIS, polarized $pp$ collisions, and all results are consistent with $|\Delta G|\le0.3$. 

The sign of $\Delta G$ is so far undetermined.  If $\Delta G$ is positive, $L_z$ is small.  If $\Delta G$ is negative, one will need large $L_z$ from quarks and gluons.  So the spin crisis remains unresolved after 20 years of experiments.

\subsection{The Challenge of $N^*$ and $\Delta$ Resonances}

Both quark model and lattice calculations predict scores of $N^*$ and $\Delta$ resonances, and most of them remain missing~(see~Fig.~5).  The claimed $N^*$ and $\Delta$ resonances remain stuck in the PDG with their poor star ratings since before 1996.  Thus, for example, of the 20 reported resonances with $M>2000$~MeV, 15 remain stuck with 1 and 2 stars, i.e., their existence is doubtful.  The old data was mostly produced with pion beams, and there are no new pion beams around,  To boot, pions can not be polarized!!

The only hope is to search for the resonances in photo-- and electro--production, and decays into final states with $\eta$, $\eta'$, $\omega$.  However, Capstick and Roberts have warned that these amplitudes tend to be ``quite small'' and the going is going to be tough.  Nevertheless, valiant groups at MAMI, ELSA, and JLab are trying.  On the analysis side, new and more comprehensive general purpose tools of PWA analysis are being developed.  It is time that these efforts had some good luck!

\begin{figure}[!tb]
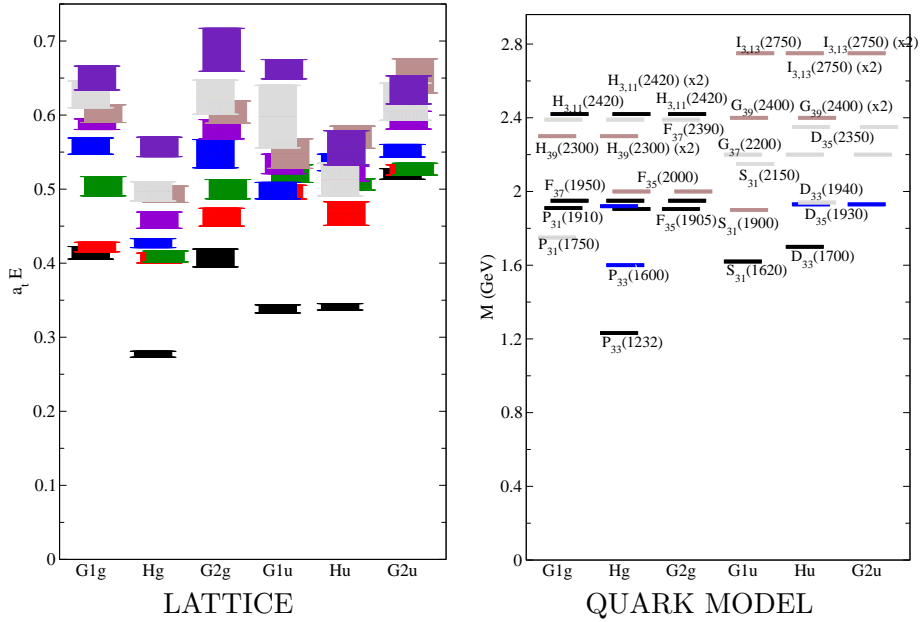

\begin{center}
\begin{tabular}{cc}
\includegraphics[width=2.3in]{hadron07_figs/DeltaMasses.eps}
&
\includegraphics[width=2.3in]{hadron07_figs/DeltaMasses_Expmnt.eps}
\\
\small LATTICE
&
\small QUARK MODEL
\\
\end{tabular}
\end{center}
\caption{Predictions for $N^*$ and $\Delta$ resonances as functions of lattice symmetry variables:  (Left) Lattice predictions, (Right) Quark model predictions.}
\end{figure}

\subsection{$\Lambda,~\Sigma$, and $\Xi$ Baryon Resonances}

The situation here is also quite bleak.  PDG07 summarizes it as follows:
\begin{quote}
$\Lambda$ and $\Sigma$: ``The field remains at a standstill and will only be revived if a kaon factory is built.''\\
$\Xi$: ``Nothing of significance on $\Xi$ resonances has been added since our 1998 review.''
\end{quote}
What can we expect in the near future?  The only kaon factory on the horizon is JPARC and hopefully they will put high priority on $\Lambda$ and $\Sigma$ \textbf{formation} experiments.  Other than that, we have only production experiments possible, in $pp$ collisions at COSY, and photoproduction experiments at JLab.  In fact, some low--lying $\Lambda$ and $\Sigma$ are being currently studied in photoproduction experiments at JLab with polarized photons, and an ambitious program of $\Xi$ spectroscopy has been proposed at JLab.  Unfortunately, we do not have any finished results so far.

\subsection{Charmed Baryons ($\bm{C=+1,~(+2),~(+3)}$)}

Here progress is more encouraging.  Adding charm quarks to the SU(3) octet and decuplet of $u,d,s$ quarks gives 18 baryons with one $c$-quark, 6 baryons with two $c$-quarks and one baryon $\Omega^{++}_{ccc}$ with three $c$-quarks.

Prior to 2005, most of the charm baryon results came from CLEO and ARGUS from $e^+e^-$ annihilations in the Upsilon$(4S)$ region, and from FOCUS at Fermilab and NA38 at CERN.  Now that the $B$--factories have weighed in, we have five new charmed baryons just this year.  BaBar has reported the discovery of $\Lambda_c(2940)$ and $\Omega_c(2770)$, and Belle has reported $\Sigma_c(2800)$, $\Xi_c(2980)$, and $\Xi_c(3080)$.  These are clean-cut states with small widths.  For example, $\Gamma(\Xi_c(3080))=6.2$~MeV!

Since 2002, we have had SELEX report the doubly charmed $\Xi^+_{cc}(3519)$, but nobody else (Belle, BaBar) seems to find any evidence for it.  So, it remains hanging.  The holy-grail particle $\Omega^{++}_{ccc}$ remains undiscovered so far.  Let me only add the hope that PANDA can reach for it.

\subsection{Bottom Baryons $\bm{(B=+1)}$}


One expects bottom baryons $\Lambda_b$, $\Xi_b$, $\Sigma_b$, and $\Omega_b$ just as the charmed baryons.  Before 2006, only one bottom baryon $\Lambda_b^0$ was known.  Now, from CDF and D\O~we have $\Sigma_b^\pm$, $\Sigma_b^*$, and $\Xi_b$.  These are extremely challenging measurements, resolving states at $\sim6$~GeV separated by $\sim20$~MeV, e.g., $m(\Sigma_b^*)-m(\Sigma_b^\pm)=21.2\pm0.2$~MeV.

\subsection{Threshold States of Two Baryons}

Long ago, in the era of prehistory, there was great excitement about the possible existence of \textbf{dibaryons}, which were predicted in bag--models.  Many, many people (including me) made many, many measurements, and in the end, nobody found any dibaryons that anybody else would believe.  Then there was the search for \textbf{baryonium}, the bound state of a baryon--antibaryon.  Again, many measurements were made searching for a $p\bar{p}$ baryonium, and finally it was agreed that there was no evidence for it.  Recently, a pseudo-baryonium has resurfaced as an enhancement in $p\bar{p}$ invariant mass at threshold, observed by BES in $J/\psi\to\gamma p\bar{p}$.  BES interpreted it as due to a below-threshold resonance with the $p\bar{p}$ bound by about 20 MeV.  Belle and BaBar also observed similar enhancements over phase space in various $B$ decays, but did not venture into the bound-state conjecture.  The enhancement was not observed in $J/\psi\to\pi^0p\bar{p}$ or $\psi'\to(\pi^0,\eta)p\bar{p}$, and the resonance interpretation has languished.  In the meanwhile, BES has reported similar near-threshold enhancements in $p\Lambda$, $\Lambda\overline{\Lambda}$, and $\omega\phi$ invariant mass.  The enhancements appear real, but the resonance interpretations appear more like wishful thinking.  It is more likely that these are manifestations of near-threshold final state interactions when the two particle go out with very small relative momenta.  Certainly more experimental and theoretical investigations are desirable, and some are in progress at CLEO.

\subsection{Baryon Summary}

Little progress has been made with light quark baryons.  Optimism for future progress has to rest particularly on what JLab and JPARC can do.  Heavy quark baryons have shown more life recently due to contributions from the B--factories at KEK and SLAC.

To cap this section, let me mention that the exotic baryons, the $\Theta^+(1540)$ pentaquark, $\Phi(1860)$, $\Theta_c^0(3100)$ appear to have mercifully expired!

\section{LIGHT QUARK MESONS}

Once again I begin with my semi-serious quotation from the Bible, the PDG.  In contrast to the baryons, which did not add a single page (148/148) between 2004/2006, the mesons showed a lot of activity, going from 358 to 430 pages.  Most of the new activity (90\% of it) came from the heavy quark (charm and bottom) sectors, with charmonium ($c\bar{c}$) showing a 63\% increase.  This indicates that a lot of the challenges in the light quark (up, down, strange) mesons have remained unanswered.  So, let me begin with them.

\begin{figure}[!tb]
\begin{center}
\begin{tabular}{ll}
\includegraphics[width=2.3in]{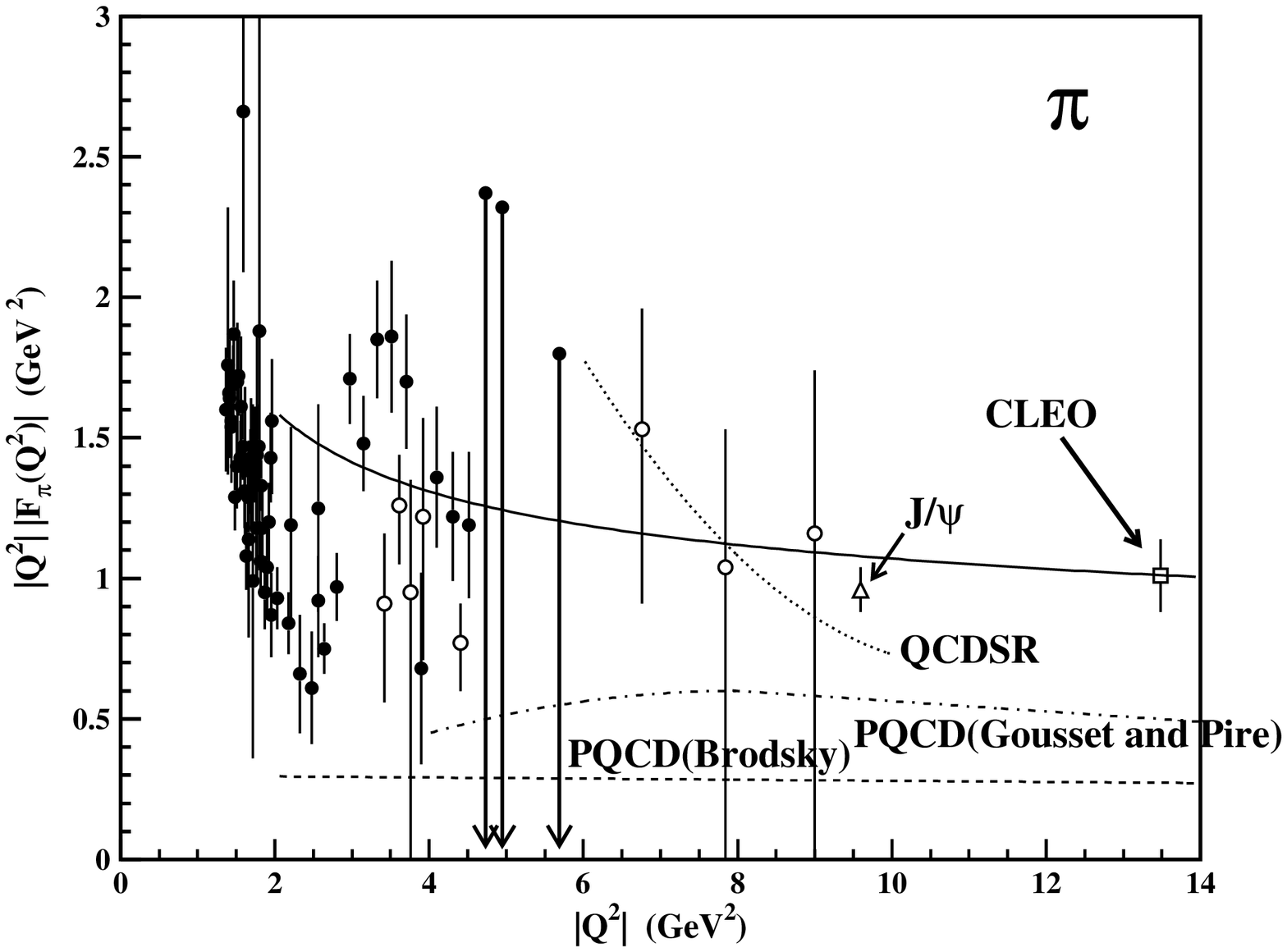}
\includegraphics[width=2.3in]{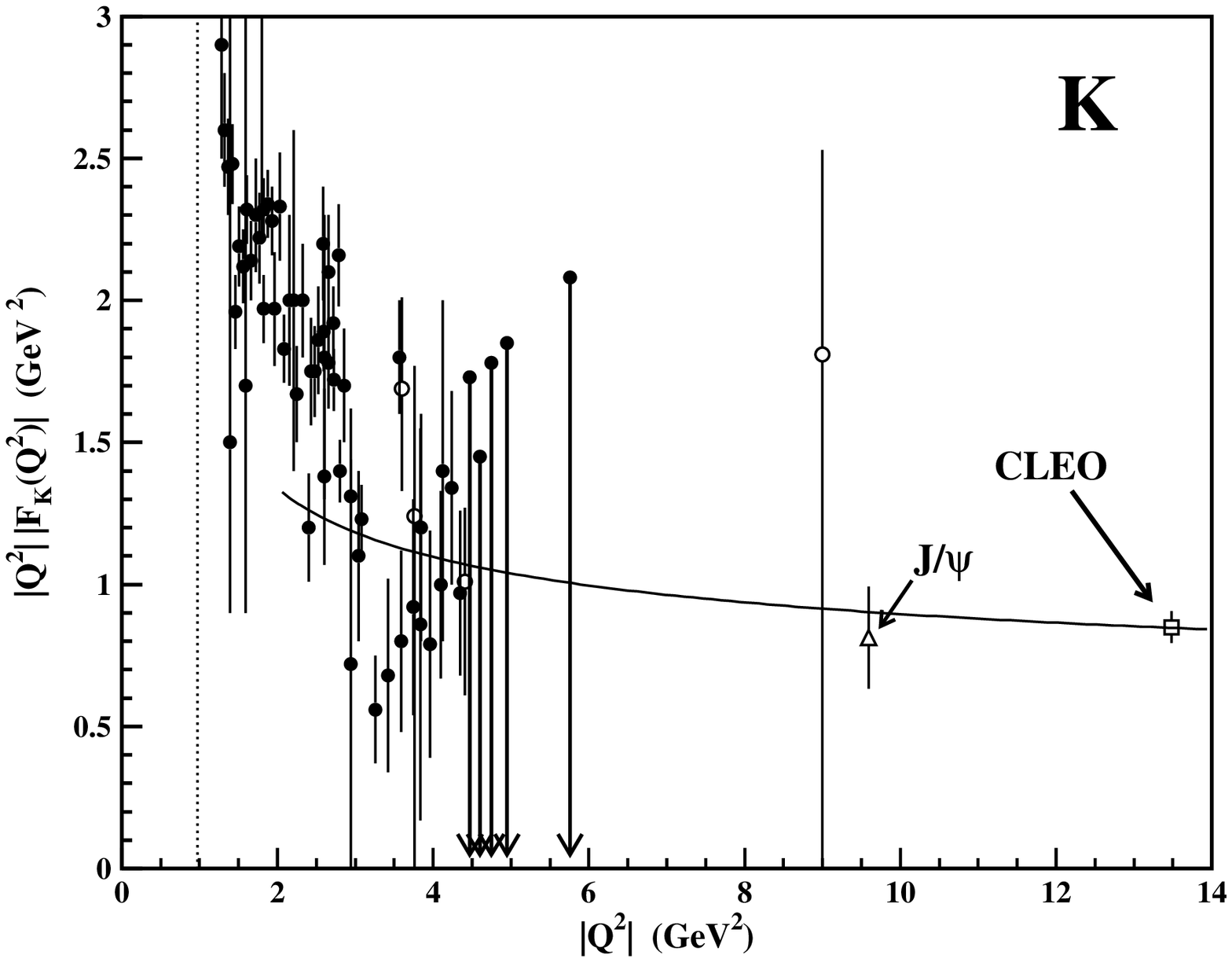}
\end{tabular}
\end{center}
\caption{Timelike form factors: (left) for pions, (right) for kaons presented as $Q^2F_m$ versus $|Q^2|$.  The solid curve, arbitrarily normalized, shows variation of $\alpha_S$.}
\end{figure}

\begin{figure}[!tb]
\begin{center}
\includegraphics[width=2.in]{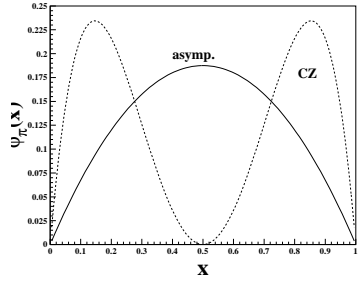}
\end{center}
\caption{Schematic illustration of the alternative momentum distributions of the quarks in a pion.  The solid curve is the asymptotic single hump distribution, and the dotted curve is the QCDSR--based dumbbell-like distribution.}
\end{figure}

\subsection{The Challenge of the Meson Form Factors}

Earlier, I posed the question, ``Is it too much to ask what the proton looks like?'' Now I am ready to ask what ought to be a simpler question because it involves only two quarks, ``What does a meson look like?''  Unfortunately, the answer to this question is even more elusive.  Some of you may recall the animated controversy between two illustrious theoretical groups --- Brodsky \& colleagues, and Isgur \& Llwellyn--Smith, about when in $Q^2$, 10~GeV$^2$ or 100's~of~GeV$^2$, pQCD begins to become valid.  And the primary experimental data available to either side (with errors less than 50 to 100\%) were pion form factors for spacelike momentum transfers $Q^2<4.5$~GeV$^2$ (see~Fig.~6).  No wonder one could not decide whether the pion distribution amplitudes looked like a dumbbell or a bell (see~Fig.~7).  Long after the original controversy, lots of theoretical predictions kept on being made, unconstrained by any new experimental data,  Well, the situation has changed, because new measurements of \textbf{pion} and \textbf{kaon} form factors for timelike momentum transfers of $Q^2=13.48$~GeV$^2$ have been made at CLEO with errors of $\pm13\%$ and $\pm5\%$, respectively.  As Figure~6~(left) for pions shows, none of the theoretical calculations, either pQCD or QCD sum rule based, make any sense at all.  This is undoubtably a strong challenge to the theorists.  For kaons there are no theoretical predictions.  The naive expectation $F(\pi)/F(K)=f^2(\pi)/f^2(K)=0.67\pm0.02$ is also found to be in disagreement with the CLEO result of $1.19\pm0.15$.  That adds to the challenge for the theorists.  There is also a challenge for the experimentalists, in this case for BES III, to measure these form factors in the $Q^2=s=4-10$~GeV region to see if $Q^2F_\pi$ varies as $\alpha_S$, as predicte by pQCD counting rules, and if the ratio $F(\pi)/F(K)$ changes.

\subsection{Light Quark Scalars}

This has been \textbf{the} hot topic in the light quark sector for a long time, and has become even more so because it intersects wth the question of the lowest mass $0^{++}$ glueball, and even with the very concept of what constitutes a ``resonance''.  It is such a hot topic that a recent review (arxiv.org/abs/0708.4016[hep-ph]), devotes 60 pages to the topic.  It offers several provocative suggestions with many of which I do not agree, but then the authors honestly admit that they offer ``a series of clear statements with little reasoning or justification.''  

The essential problem with the scalars is that in the quark model, with three light quarks you expect three scalars, two isospin--zero $f_0$ and one isospin--one $a_0$.  But we have an embarressment of riches.  We have at least five $f_0$'s: $f_0(600)$ or $\sigma(600)$, $f_0(980)$, $f_0(1370)$, $f_0(1500)$, and $f_0(1710)$.  So, we have to somehow disqualify at least three of these as non--$q\bar{q}$ mesons.

\subsubsection{The Challenge of the $\sigma$ and $\kappa$}

Even at HADRON05, the reality of the $\sigma$ meson which decays almost exclusively into $\pi\pi$ was open to question.  Now, all the evidence has converged, and there appears to be little disagreement with the conclusion that $\sigma$ is a real Breit-Wigner resonance with it proper pole structure, and (give or take 10 MeV or so)
$$M(\sigma)=480~\mathrm{MeV},\quad\Gamma(\sigma)=570~\mathrm{MeV}$$
What is still an open question is ``what is $\sigma$?''  The debate is wide open.  Is it $q\bar{q}$?  Is it a glueball?  Is it a 4--quark state?  Or, is it (despite its Breit--Wigner character) the result of a final state $\pi\pi$ interaction?  The challenge here is to find a way to distinguish between these various possibilities.  I cannot think of anything except that a strong production of $\sigma$ in two photon fusion would perhaps eliminate the glueball hypothesis.

The case of $\kappa(980)$ is more complicated and controversial.  The ``evidence'' comes from $K\pi$ scattering (LASS), $K\pi$ production in decays of $D$ mesons (FOCUS), and radiative decay of $J/\psi$ (BES).  The different analyses give very different masses
$$M(\kappa)=658-841~\mathrm{MeV}$$
albeit with large errors.  The spread in widths is even worse
$$\Gamma(\kappa)=410-840~\mathrm{MeV}$$
In my mind, the existence of $\kappa$ remains questionable, although I hold in high respect the work based on dispersion relation based analysis of $K\pi$ scattering.  Personally, I intend to look at our own essentially background free data for $D\to K\pi\pi$ to see if we can shed some new light on both $\sigma$ and $\kappa$.

\subsubsection{The $f_0(980),~a_0(980)$, and 4-quark States}

Unlike the $\sigma$ and $\kappa$, there is no doubt about the existence of $f_0(980)$ and $a_0(980)$.  They have been observed in $e^+e^-$ and $p\bar{p}$ annihilation, in $pp$ central collisions, in two photon fusion, and in radiative decays of $J/\psi$ and $\phi$ (for which we have an excellent contribution from KLOE here).  There is no doubt that $f_0$ and $a_0$ are relatively narrow ($\Gamma<100$~MeV) and have strong decays to $K\overline{K}$.

The fact that the masses of $f_0$ and $a_0$ are very close, $M(K^+K^-)=987.5$~MeV and $M(K_SK_L)=995.3$~MeV, and that they decay strongly into $K\overline{K}$, has given rise to the long--standing proposition that these are $K\overline{K}$ molecules.  On the other hand, a canonical calculation of $q\bar{q}$ masses by Godfrey and Isgur predicts the first $f_0$ and $a_0$ to each have masses of 1090~MeV, within shooting distance of 980~MeV. Unfortunately, Godfrey and Isgur also predict much larger ($\times5$) total widths.  Not knowing any better, I am ready to consider the wave functions for these states mixtures of $q\bar{q}$ and four quark ($qq\bar{q}\bar{q}$ or $q\bar{q}q\bar{q}$) configurations.  Unless somebody can devise a ``smoking gun'' measurement which would determine which configuration (if any) is dominant, I am content to live with this ad-hoc compromise.

I must emphasize, however, that even these mixed configuration mesons must be included in the $q\bar{q}$ meson count.  Although many more 4-quark states can be configured, it is generally agreed that only those 4-quark configurations survive which can mix with $q\bar{q}$.  As long as we are in the land of unproven conjectures, it is my conjecture that the $f_0(q\bar{q})$ and $a_0(q\bar{q})$ predicted by Godfrey \& Isgur at 1090 MeV have moved down by mixing with four quark configurations, or something else to, 980 MeV, and $f_0(980)$ and $a_0(980)$ are the legitimate members of the $q\bar{q}$ scalar nonet.  This is, of course, in contradiction to what the review authors of the PDG would have me believe.  They propose an inverted spin--orbit splitting with $f_0$ and $a_0$ nearly 110 MeV \textbf{above} $f_2$ and $a_2$!!

\subsubsection{The $f_0(1370),~f_0(1500),~f_0(1710)$ and the Glueballs}

Sometime ago there were questions about this triad.  Does $f_0(1370)$ really exist?  Does $f_0(1710)$ really have $J^{PC}=0^{++}$?  There is now widespread belief that $f_0(1370)$ exists, and it is firmly established that $f_0(1710)$ has $J^{PC}=0^{++}$.  The challenge now is:  can we draw any conclusions about \textbf{the $\bm{0^{++}}$ scalar glueball?}  Ten years ago, there were almost partisan discussions about which one of these is \textbf{THE (pure) GLUEBALL}, and all kinds of ``semi-smoking gun'' criteria were suggested to make the choice.  Among them were:  glueballs should be narrow (why? and how narrow?), glueballs should decay flavour-blind, glueballs should be supernumary to quark model expectations.  Now everybody agrees that the scalar glueball not only can, but must mix with all of the other $f_0$'s in its neighborhood.  So, the ``smoking gun'' does not have to smoke very much!   The mixed glueballs can be broad, and the their decays can have large departures from flavor-blindness.  \textbf{In fact, the search for the uniquely identifiable glueball does not make much sense!}

At the generic level, we have expectations of four low mass isoscalars, two $q\bar{q}$, one glueball, and throw in a four-quark state.  And they all mix, to give us $f_0(980)$, $f_0(1370)$, $f_0(1500)$, and $f_0(1710)$.  Problem solved, or is it?  Let the games go on!  And isn't that challenging!!

\textbf{The $\bm{2^{++}}$ tensor glueball} is likely to have the same fate as the scalar.  The narrow $\xi(2230)$ has evaporated, Godfrey and Isgur predict \textbf{six} $f_2(q\bar{q})$ below 2400~MeV, and at least \textbf{twelve} have been claimed by one experiment or another.  So the putative tensor glueball will also have plenty of friends to mix with!

As long as we are talking about glueball admixtures in mesons, it is worth noting that a beautiful measurement at KLOE has estimated the gluonium content of $\eta'(958)$ to be $14\pm4\%$, assuming that $\eta(548)$ has none.  They do it by a long awaited precision measurement of $\mathcal{B}(\phi\to\eta'\gamma)/\mathcal{B}(\phi\to\eta\gamma)=(4.77\pm0.21)\times10^{-3}$.


\subsubsection{Light Quark Hybrids}

As is well known, three $J^{PC}=1^{-+}$ states, $\pi_1(1400,~1600,$~and~$2000)$ have been reported by the Brookhaven and Protvino groups.  Since $J^{PC}=1^{-+}$ is forbidden for $q\bar{q}$ mesons, these states are obviously ``exotic''.  However, their interpretation as $q\bar{q}g$ hybrids is not universally accepted.  This status remains unchanged since their discovery. It is claimed that photoproduction of hybrids holds great promise, particularly at JLab, but that is also not without controversy.  Unfortunately, experimental resolution of this controversy has to wait for the JLab upgrade, which may come as late as ten years from now.

\section{HEAVY QUARK MESONS}

Heavy quark (charm, beauty or bottom) mesons have several advantages over their light quark partners.  They do not have the multitude of light quark mesons in their neighborhood.  So their spectra are generally cleaner.  Also, because $\alpha_S$ at heavy quark masses is smaller, and relativistic effects are weaker, perturbative predictions for heavy quark mesons are expected to be more reliable.

\subsection{The Challenge of the Open Flavor Mesons}

The open flavour mesons make the heavy-light system ($Q\bar{q}$, $\overline{Q}q$).  Life is supposed to become simpler because the heavy quark $Q$ ($c$ or $b$), with spin $S_Q$, provides a more-or-less static core around which the light quark ($u,d,s$) with $j_q=l+S_q$ orbits.  This gives rise to the heavy-quark effective theory, or HQET, for the heavy-light system with $J=j_q+S_Q$, which has been very successful.

\subsubsection{The Open Charm or $D$ Mesons (=$c\bar{n},~c\bar{s}$)}

In 2003, BaBar and CLEO discovered $D_s^*(c\bar{s})$, $J=0^+$ and $1^+$ mesons which were expected to lie above the thresholds for $DK$~(2367~MeV) and $DK^*$~(2508~MeV) and therefore to be wide.  Instead, they turned out to have masses 2318~MeV and 2456~MeV, i.e., each 50 MeV below their respective thresholds, and both were $<5$~MeV wide.  As always, when the unexpected happens, there is no dearth of possible explanations for the observed mesons, $c\bar{s}$ displaced by mixing, $DK$ molecules, tetraquark, etc., but there is no consensus.  In the meanwhile, there are more challenges.

The analogues of $D_s^*(0^+,1^+)$, the $D^*(\sim2218,0^+)$ and $D^*(\sim2360,1^+)$ which are expected to be broad have not yet been identified.  Further, BaBar has announced a new relatively narrow $D_s$ with $M/\Gamma=2857/48$~MeV, and a broad $D_s$ with $M/\Gamma=2688/112$~MeV.  Belle can not find $D_s(2857)$ and reports a $J^P=1^-$~$D_s(2708)$, which is presumably the same as BaBar's, $D(2688)$.  Are these radially excited $D_s$ states?  Time will tell, as more radially excited states are discovered.

\subsubsection{The Open Beauty or $B$ Mesons (=$b\bar{n},~b\bar{s},~b\bar{c}$)}

This is the domain of CDF and D\O~contributions, and they have made many precision measurements of $B$-mesons, ($B_1^0,~B_{s1}^0$), ($B_2^0,~B_{s2}^0$).  The latest triumph is a precision measurement of the $B_c$ meson, $M(B_c)=6274.1\pm4.1$~MeV.  

A remarkable, at least to me, conclusion is that the $B_s$ mesons are always $100\pm5$~MeV heavier than the $B$ mesons.  This is exactly what was observed for the $D$-mesons.  It looks like that the $s$-quark marble is just 100~MeV heavier than the $u,d$ quark marbles.  Life should always be so simple!

Of course, the main thrust in the study of the open flavor mesons is weak interactions, decay constants, form factors, and CKM matrix elements, and the Standard Model.  CLEO and CDF,  D\O, Belle, and BaBar have been working hard on these measurements, and comparison with lattice predictions, but I will continue to confine myself to strong interactions.

Let me now turn to Quarkonia, $c\bar{c}$ charmonium, $b\bar{b}$ bottomonium, and the newly discovered surprising states.

\subsection{Quarkonia: The Hidden--Flavour Mesons}

The $SU(3)$ light quarks have such similar masses (within 100~MeV) that it is difficult, and even meaningless, to look for pure $u\bar{u},~d\bar{d},~s\bar{s}$ mesons.  They invariably mix (despite the near purity of $\phi$ as a $s\bar{s}$).  However, the charm quark and the bottom quark have much different masses and essentially do not mix with other flavors.  So we have pure $c\bar{c}$~charmonium and $b\bar{b}$~bottomonium.  By far the greatest activity in strong interaction physics has been in the charmonium region, which I define as $\sim3-5$~GeV.  So let me begin with charmonium and what has come to be known as charmonium-like mesons.

\subsubsection{Challenges in Charmonium Spectroscopy}

We are all familiar with the story of the discovery of $J/\psi$ and the beginning of the QCD era with it.  Over the years tremendous activity followed at SLAC, Frascati, DESY, ORSAY, and more recently at Fermilab, CLEO, and BES in laying down the QCD--based foundation of quarkonium spectroscopy.  BES and CLEO have in recent years made many high precision measurements of decays of bound charmonium states, but here I want to talk about several recent discoveries.  \vspace{8pt}

\noindent\textbf{(a) The Spin--Singlet States}\vspace{6pt}

A close examination of the spectroscopy of charmonium states will reveal that most of what was discovered and studied until recently was about \textbf{spin--triplet states}, the $\psi(^3S_1)$, $\chi_{cJ}(^3P_J)$ states of charmonium and $\Upsilon(^3S_1)$, $\chi_{bJ}(^3P_1)$ states of bottomonium.  The \textbf{spin--singlet states} were too difficult to access, and remained unidentified (with the exception of $\eta_c(1^1S_0)$).  This meant that we had very little knowledge of the hyperfine interaction which splits the spin--singlet and spin--triplet states.

To emphasize the importance of the spin--spin, or hyperfine interaction, let me remind you (as Prof. Miani also did) of the textbook discussion of the ground state meson masses in the elementary quark model, i.e.,
$$M(q_1\bar{q}_2)=m(q_1)+m(q_2)+\frac{8\pi\alpha_S}{9m_1m_2}|\psi(0)|^2\vec{\sigma}_1\cdot\vec{\sigma}_2$$
In other words, the only ingredient required other than the quark masses is the spin--spin, $\vec{\sigma}_1\cdot\vec{\sigma}_2$ interaction.  It is, of course, the same interaction which gives rise to the hyperfine, or spin--singlet/triplet splitting in quarkonium spectra.

Yet, until very recently, all that we knew was the singlet--triplet splitting for $\eta_c(1^1S_0)$ and $J/\psi(1^3S_1)$, with $\Delta M_{hf}(1S)=117\pm2$~MeV.  We knew nothing about whether the hyperfine interaction varies with the radial quantum number or quark mass, or what all of it means with respect to the spin dependence of the long range $q\bar{q}$ interaction which is dominated by its confinement part.\vspace{8pt}

\noindent\textbf{\underline{The $\eta_c(2^1S_0)$ State}}\vspace{6pt}

The breakthrough came in 2003 with the identification of $\eta_c'(2^1S_0)$ by Belle, CLEO, and BaBar.  The result, $\Delta M_{hf}(2S)=48\pm2$~MeV, nearly 1/3 of $\Delta M_{hf}(1S)$ came as a surprise.  Although one or another potential model calculator will tell you that this was no surprise, the fact is that they were fully at peace with the old (wrong by $\sim$ factor two) Crystal Ball value, $\Delta M_{hf}(2S)=92\pm5$~MeV.  The most common explanation offered for the present result is that it is due to mixing with the continuum states, but in my mind a still-open possibility is the existence of a long--range spin--spin interaction in the confinement region.\vspace{8pt}

\begin{figure}[!tb]
\begin{center}
\includegraphics[width=3.5in]{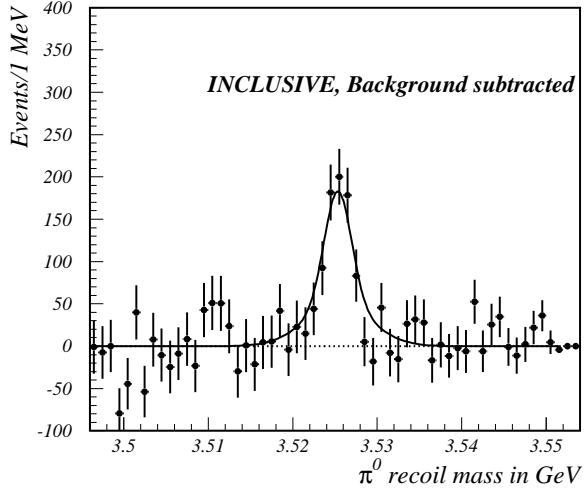}
\end{center}
\caption{Background subtracted CLEO spectrum for $h_c$ for the inclusive analysis of 24 million $\psi(2S)$.}
\end{figure}

\noindent\textbf{\underline{The $h_c(1^1P_1)$ State}}\vspace{6pt}

The second breakthrough in the understanding of the hyperfine interaction comes from the even more recent identification of the $P$--wave singlet state $h_c(1^1P_1)$, which had eluded numerous earlier attempts. Two years ago, CLEO announced the discovery of $h_c$ in both inclusive and exclusive analysis of the isospin--forbidden reaction
$$\psi(2S)\to\pi^0h_c,~h_c\to\gamma\eta_c,~\pi^0\to2\gamma$$
The data, based on $\sim3$~million~$\psi(2S)$, had limited statistical precision, as did a recent E835 attempt.   Now CLEO has analyzed their latest data for 24~million~$\psi(2S)$.  More than a thousand $h_c$ have been identified, illustrating the adage that yesterday's ``enhancement'' can become today's full--blown resonance (see Fig.~8). The precision result obtained by CLEO is 
$$M(h_c)=3525.34\pm0.19\pm0.14~\mathrm{MeV,}$$ 
which leads to 
$$\Delta M_{hf}(1P)\equiv M(\left<^3P_J\right>) - M(^1P_1)=-0.04\pm0.19\pm0.15~\mathrm{MeV.}$$
 This would appear to be just what one expected, because the one-gluon exchange hyperfine interaction is supposed to be a contact interaction and therefore non-existent in $L\ne0$~states, i.e., $\Delta M_{hf}(1P)=0$.  Actually, the above experimental result is based on determining 
$$M(^3P_J)=M|\left<^3P_J\right>|=(5M(^3P_2)+3M(^3P_1)+M(^3P_0))/9.$$  
J.~M.~Richard and A.~Martin have repeatedly pointed out that this is the wrong way to determine $M(^3P_J)$, because the $L\cdot S$ splitting of $^3P_J$ states is not perturbatively small, being 141~MeV.  The correct way to determine $M(^3P_J)$ and $\Delta M_{hf}(1P)$ is to turn the $L\cdot S$ and $T$ interactions off and directly determine $M(^3P_J)$ and $M(^1P_1)$.  In fact, when this is done in a typical potential model calculation with no explicit long-range hyperfine interaction, $\Delta M_{hf}=9$~MeV is obtained.  So how can we explain the $\Delta M_{hf}(1P)=0$ experimental result?  Apparently, there are subtle problems connected with the regularization of the spin--dependent interactions, and nobody really knows how to handle these subtleties.

In any case, with $\eta_c'$ and $h_c$ identified, the spectrum of the bound states of charmonium is now complete, and we can move on to the unbound states above the $D\overline{D}$ threshold at 3739~MeV.\vspace{10pt}

\noindent\textbf{(b) Charmonium-like States, or The Bounty of Unexpected States Above $D\overline{D}$}\vspace{6pt}

The vector states $\psi(3770,4040,4160,4415)$ above the $D\overline{D}$ threshold at 3.74~GeV have been known for a long time.  However, little more than their total and leptonic widths was known.  Now we know a lot more.  CLEO and BES, and more recently Belle, have contributed much new information about their decays into $D\overline{D}$, $D\overline{D}_s$, $D_s\overline{D}_s$.  The CLEO work is primarily motivated by trying to find the optimum energies to run in order to produce maximum yields of $D$ and $D_s$ for weak interaction studies.

The real excitement in this domain of spectroscopy has come about by the discovery of seveeral unexpected states by the $B$-factories of Belle and BaBar.  It began with X(3872).  Then came the states X,~Y,~Z with nearly degenerate masses of 3940~MeV.  This was followed by Y(4260).  And now we have reports of Y(4360), Y(4660), X(4160), and Z$^\pm$(4433).

All these states decay into final states containing a $c$ quark and a $\bar{c}$ quark, hence the designation ``charmonium-like''.  Another point worth noticing is that while X(3872) and Y(4260) have been observed by several laboratories, and X(4360) perhaps by both Belle and BaBar, the  X,~Y,~Z(3940), X(4160), and X(4660) have been only reported by Belle, with an ominous silence by BaBar.  And finally, even a charged state $Z^\pm(4433)$ has just been claimed by Belle!\vspace{30pt}

\noindent\textbf{\underline{The Challenge of X(3872)}}\vspace{6pt}

Of all the unexpected new states, only X(3872) is firmly established as a single narrow resonance with
$$M(\mathrm{X}(3872))=3871.4\pm0.6~\mathrm{MeV}, \quad \Gamma(\mathrm{X})<2.3~\mathrm{MeV}$$
From the beautiful angular correlation studies done by CDF, its spin is limited to $J^{PC}=1^{++}$ or $2^{-+}$.  The discovery mode of its decay was X(3872)$\to\pi^+\pi^-J/\psi$, but many other modes have been studied since.  There were originally many theoretical suggestions for the nature of X(3872), but the limited choice of spin now only allows $1^{++}~\chi_{c1}'(2^3P_1)$ or $2^{-+}(1^1D_2)$ in the charmonium option, and $1^{++}$ in the popular $D\overline{D}^*$ molecule or tetraquark options, which were among the first and almost obvious suggestions made because of the very close proximity of $M($X$(3872))$ to the sum of $D^0$ and $D^{*0}$ masses.  A recent precision measurement of the $D^0$ mass makes the binding energy of the molecule very small, $0.6\pm0.6$~MeV, which has a strong bearing on the $D^0$ and $\overline{D}^{*0}(\to\overline{D}^0\pi^0)$ falling apart.  The challenge for the experimentalists is obviously to measure both $M($X$(3872))$ and $M(D^0)$ with even greater precision, so that even stricter limits  on the $D^0\overline{D}^{0*}$ binding energy can be put.  Also, Belle needs to measure the decay $\mathrm{X}\to D^0\overline{D}^0\pi^0$ with greater precision, because their present measurement is at strong odds with the prediction of the molecule model.  Recently, the branching factor for X(3872) decaying to $\overline{D}^{*0}D^0$ has been reported.  In order to explain its nearly factor 200 larger value than can be accomodated in the molecular model, it has been claimed that there is another resonance just a few MeV away.  However, the experimental evidence for this is very shaky.

So what is X(3872)?  I consider the question still open.\vspace{8pt}

\noindent\textbf{\underline{The Challenge of Y(4260)}}\vspace{6pt}

The Y(4260) has been observed by BaBar, CLEO, and Belle in ISR production and decay into $\pi\pi J/\psi$.  The production in ISR ensures that its spin is $1^{--}$.  The fact that its mass is precisely where $R\equiv\sigma(e^+e^-\to\mathrm{hadrons})/\sigma(e^+e^-\to\mu^+\mu^-)$ has a deep minimum indicates that it is a very unusual vector.  Also, all the charmonium vectors up to 4.4~GeV are spoken for.  These problems with a charmonium interpretation have led to the suggestion that Y(4260) is a hybrid $1^{--}$, and you have already heard impassioned advocacy of it.  So I will stay away from it, except to point out that if this is true, we should expect to see $1^{-+}$ and $0^{-+}$ hybrids at nearby lower masses.

On the experimental side, new problems have emerged.  Belle has revived the question of whether Y(4260) is a single resonance or two, Y(4008) and Y$'$(4247).  Not only that, Belle also reports that the peak positions of Y are different in its decays to $\pi^+\pi^-J/\psi$ and $\pi^+\pi^-\psi(2S)$ by $\sim120$~MeV.  So, what looked like a simple state, perhaps hybrid, now appears to be rather complicated, and the rush to judgement about its nature might be premature.\vspace{24pt}

\noindent\textbf{\underline{The Challenge of X, Y, Z(3940)}}\vspace{6pt}

In quick succession Belle reported three different states produced in different initial channels, and decaying into different final states, but all having nearly identical masses.  I will not go into the details which you have heard in several plenary and parallel talks, but will summarize the results in Table~I.

My personal summary of the situation is that X and Z exist and their charmonium interpretation requires confirmation.   I have serious doubts about Y.  In fact, BaBar's recent attempt to confirm it leads to quite different parameters.\vspace{8pt}

\begin{table}[!tb]
\begin{scriptsize}
\begin{center}
\begin{tabular}{ccc}
\includegraphics[width=1.3in]{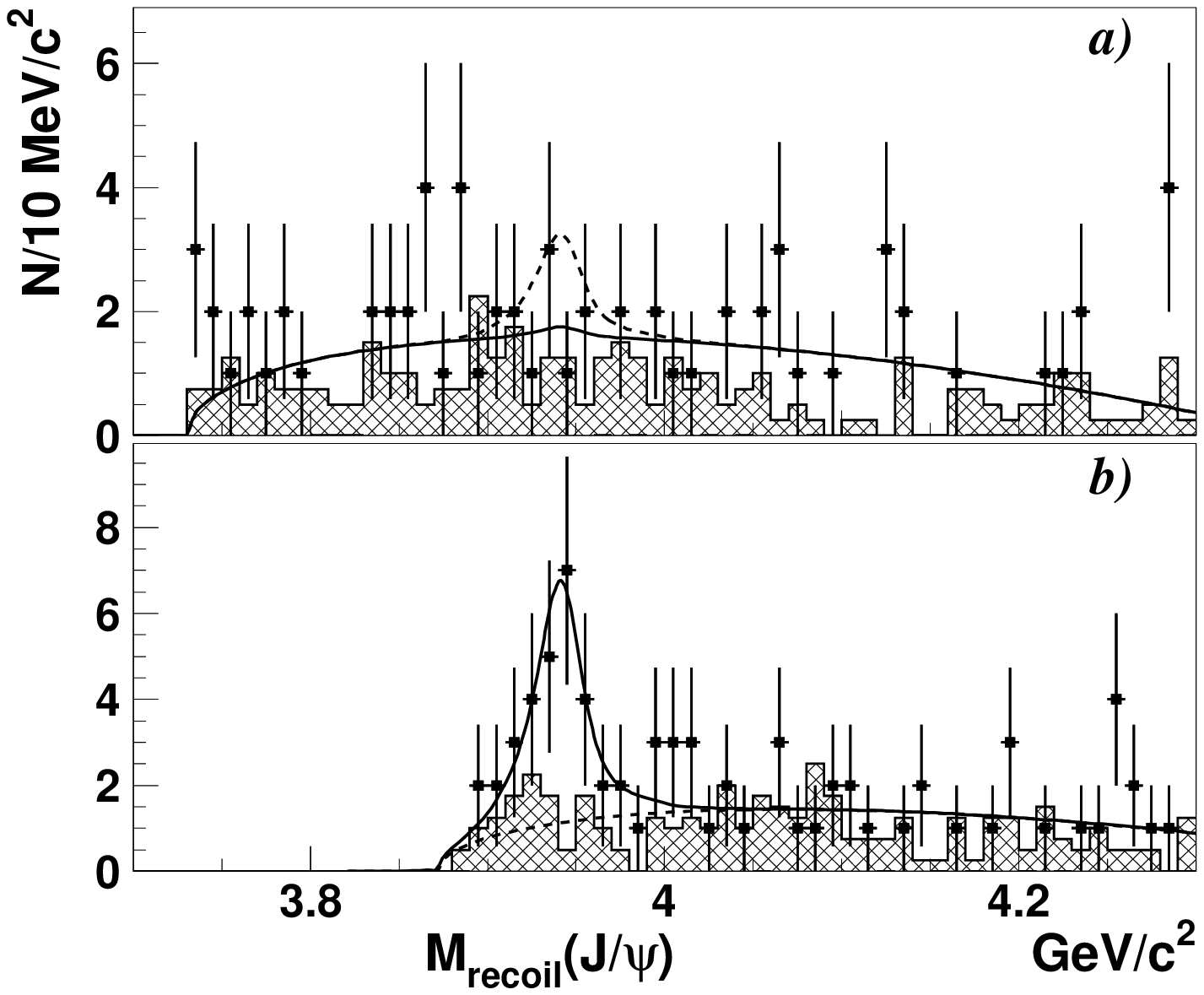}
&
\includegraphics[width=1.2in]{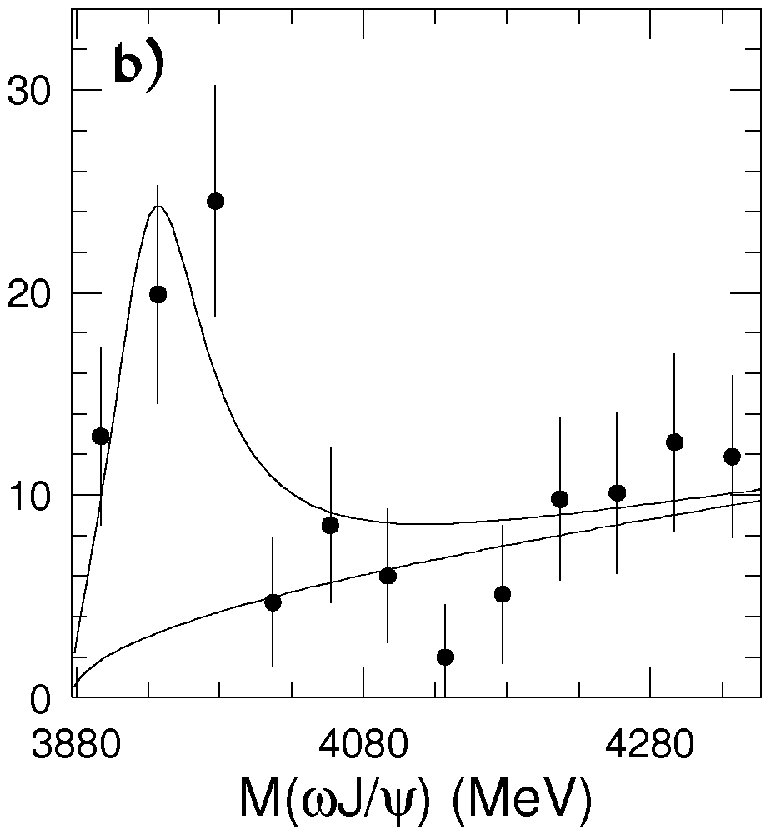}
&
\includegraphics[width=1.2in]{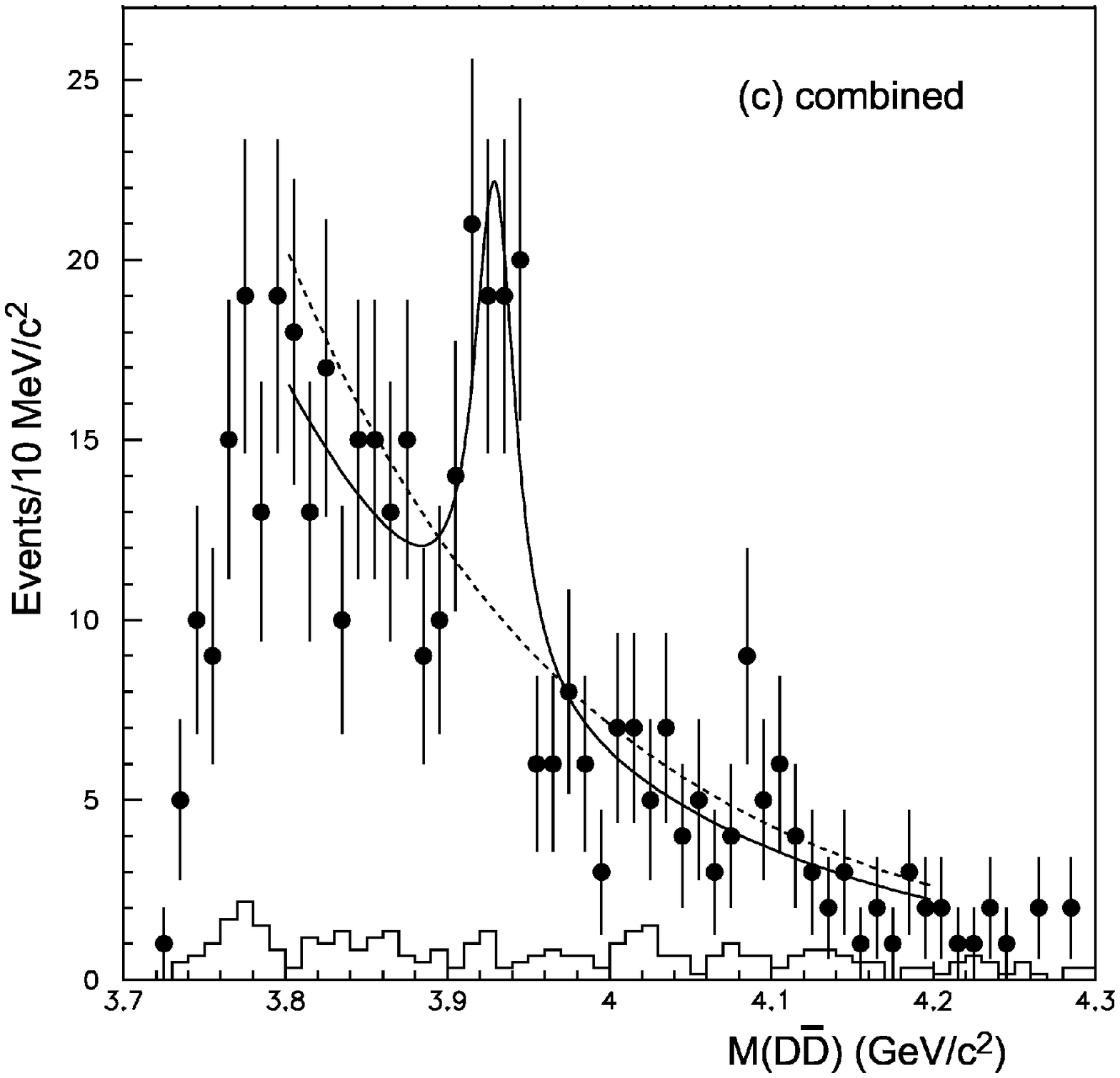}
\\
\textbf{X(3943)} & \textbf{Y(3943)} & \textbf{Z(3929)} \\
$N(\mathrm{X})=24.5\pm$ & $N(\mathrm{Y})=58\pm11$ & $N(\mathrm{X})=64\pm18$   \\
$M(\mathrm{X})=3943\pm10$~MeV & $M(\mathrm{Y})=3943\pm17$~MeV & $M(\mathrm{Z})=3929\pm5$~MeV \\
$\Gamma(\mathrm{X})=15.4\pm10.1$~MeV & $\Gamma(\mathrm{Y})=87\pm26$~MeV & $\Gamma(\mathrm{Z})=20\pm8$~MeV \\ 
\hline
\multicolumn{3}{c}{\textbf{Production}}  \\
Double Charmonium ($J=0?$) & $B\to KY$ & $\gamma\gamma$ fusion $J=2$ \\ 
\hline
\multicolumn{3}{c}{\textbf{Decay}} \\
$X\to D^*\overline{D}>45\%$ & $Y\to\omega J/\psi$ & $Z\to D\overline{D}$ \\
$X\nrightarrow D\overline{D}<41\%$, $X\nrightarrow \omega J/\psi<26\%$ & $Y\nrightarrow D\overline{D}$ & \\
\hline
\multicolumn{3}{c}{\textbf{Best Guess}} \\
$\eta_c''(3^1S_0)$ & hybrid?? & $\chi_{c2}'(2^3P_2)$ \\
\hline
\multicolumn{3}{c}{\textbf{Search for}} \\
 production in $\gamma\gamma$ & decay in $D\overline{D},~D^*\overline{D}$ & decay in $D^*\overline{D}$ \\
\hline
\end{tabular}
\end{center}
\end{scriptsize}
\caption{Spectra for X, Y, Z as observed by Belle.  Details are listed below.}
\end{table}

\noindent\textbf{\underline{Ever More States}}\vspace{6pt}

And now we have four more states.  We are running out of alphabets now. These are states at 4160, 4360, and 4664~MeV decaying into $\psi(2S)\pi^+\pi^-$, and at 4433~MeV decaying into $\psi(2S)\pi^\pm$, as listed in Table II.  BaBar does not confirm any of these.

I have to admit that the proliferation of these states is getting to be so much that one cannot help becoming incredulous.  Will all these bumps survive?  Unfortunately, yes!  For no reason other than the fact that no other measurements appear to be possible in the near future to check them.

\begin{table}[!tb]
\begin{center}
\begin{tabular}{lccccc}
\hline
 & Source & Mass (MeV) & Width (MeV) & Events & Reaction \\
\hline
X$'$ & Belle & 4160(30) & 139($^{113}_{65}$) & 24($^{12}_{8}$) & $e^+e^-\to J/\psi + D^*D^*$ \\
X$''$ & Belle & 4360(13) & 74(18) & $\sim50$ & $e^+e^-\to\psi(2S)\pi^+\pi^-$ \\
X$'''$ & Belle & 4664(12) & 48(15) & $\sim36$ &  $e^+e^-\to\psi(2S)\pi^+\pi^-$ \\
Z$^\pm$ & Belle & 4433(4) & 45($^{35}_{18}$) & 121(30) & $B\to (K)\psi(2S)\pi^\pm$ \\
\hline

\end{tabular}
\end{center}
\caption{The new states announced by Belle.}
\end{table}

\subsubsection{Challenges in Bottomonium Spectroscopy}

In principle, the bottomonium system can lead to clearer insight into the onium spectroscopy, both because $\alpha_S$ is smaller ($\alpha_S\approx0.2$) than for charmonium ($\alpha_S\approx0.35$), and also because relativistic effects are smaller.  However, $b\bar{b}$ cross sections are smaller, the states are denser, and no $p\bar{p}$ production has so far been available.  For the Upsilon ($1^{--}$) states, all we known is their masses, total widths, and branching fractions for leptonic, radiative, and $\Upsilon(nS)\to\pi^+\pi^-\Upsilon(n'S)$ decays.  A scarce $\Upsilon(3S)\to\omega\chi_{b}(2S)$ transition has been observed, but huge gaps remain.  By far the greatest gap is once again about the lack of any knowledge of singlet states.  Even the ground state of bottomonium ($\eta_b(1^1S_0)$) has never been identified, and neither has $h_b(1^1P_1)$.  Since nobody is presently planning $e^+e^-$ or $p\bar{p}$ annihilations in new searches for $\eta_b$, the only possible source is CLEO, which has the largest samples of $\Upsilon(1S,2S,3S)$ data from its earlier runs.  Indeed, serious efforts are being presently made at CLEO to identify $\eta_b$ in the radiative decay of $\Upsilon(1S)$.

This then is a great challenge --- find $\eta_b$ and $h_b$.

Of course, I have long had a dream of doing $b\bar{b}$ spectroscopy in $p\bar{p}$ annihilation.  Unfortunately, neither a fixed target $p\bar{p}$ facility (needs about 50~GeV $\bar{p}$ beams), nor a $p\bar{p}$ collider (with $\sim6$~GeV beams) appears to be on the horizon.  So, this dream is not likely to be fulfilled anytime soon.

In the meantime, one long-cherished dream may come to fruition soon if CLEO is successful in identifying $\eta_b$.

\section{Mesons in the Nuclear Medium}

It has been conjectured for a long time that meson properties, notably their masses, widths, and elementary cross sections, should be modified in the nuclear medium.  It has been predicted that the masses may change by tens of MeV, and widths may be broadened by large amounts.  Also, cross sections for meson+A collisions should be quite different than meson+$p$ collisions (color transparency).  We are now beginning to get some answers, and as is always true, more questions.  

It is claimed that color transparency has been experimentally observed at high energies, with some unexplained observations at lower energies.  The interesting problem of $J/\psi$ attenuation in heavy ion collisions, so important for the QGP question, remains provocatively open, because $\sigma(J/\psi-\mathrm{nucleon})$ in nuclear medium remains unmeasured.  

About mesons masses there are experimental controversies, for example, there are reports of a large shift in vector meson masses by KEK, and there are reports of almost no shift by JLab.  The situation at the moment is fluid, and it calls for more measurements with high precision and high mass resolution.

To summarize my own talk, let me say that many, many extremely interesting questions in hadron spectroscopy remain unanswered at present.  However, there is every hope that the upcoming facilities, PANDA at GSI, JPARC at KEK, and the 12~GeV upgrade at JLab, will rise to meet the challenges posed by these questions and the theorists will find them deserving of serious attention even in this era of the Higgs and Beyond the Standard Model!!

\section*{Epilogue}

Since this is the last talk of the Conference, let me take the opportunity, on behalf of all of us, to thank the organizers for their warm welcome, a very pleasant and successful conference, and also for the beautiful Frascati weather!

\clearpage


\end{document}